# PROTEIN THREADING BASED ON NONLINEAR INTEGER PROGRAMMING

Wajeb Gharib Gharibi
Department of Computer Engineering & Networks
Jazan University
Jazan 82822-6694, Saudi Arabia
Gharibi@jazanu.edu.sa

Marwah Mohammed Bakri
Department of Biology
Jazan University
Jazan 755, Saudi Arabia
Marwah890@gmail.com

**Abstract**

Protein threading is a method of computational protein structure prediction used for protein sequences which have the same fold as proteins of known structures but do not have homologous proteins with known structure. The most popular algorithm is based on linear integer programming.

In this paper, we consider methods based on nonlinear integer programming. Actually, the existing linear integer programming is directly linearized from the original quadratic integer programming. We then develop corresponding efficient algorithms.

## 1 INTRODUCTION

Protein structure prediction from amino acid sequence is a fundamental scientific problem and it is regarded as a grand challenge in computational biology and chemistry.

Protein threading problem also referred as the holy grail of molecular biology on the second half of the genetic code is to determine the three-dimensional folded shape (protein structure prediction) of a protein (sequence of characters drawn from an alphabet of 20 letters). It is important because the biological function of proteins underlies all life, their function is determined by their three-dimensional shape, and their shape determined by one-dimensional sequence.

The prediction is made by "threading" (i.e. placing, aligning) each amino acid contained in the target sequence to a position in the template structure, and evaluating how well the target fits the template. After the best-fit template is selected, the structural model of the sequence is built based on the alignment with the chosen template. The protein threading method is based on two basic observations. One is that the number of different folds in nature is fairly small (approximately 1000), and the other is that according to the statistics of the Protein Data Bank (PDB), 90% of the new structures submitted to PDB in the past three years have similar structural folds to the ones in PDB.

A general paradigm of protein threading consists of the following four steps: the construction of a structure template database, the design of the scoring function, threading alignment and threading prediction. The third step is one of the major tasks of all threading-based structure prediction programs, which mainly dedicated to solving the optimal alignment problem derived from a scoring function considering pairwise contacts.

As a formal presentation of the problem, let C called core be a set of m items $S_i$, called segments of length $l_i$. This set must be aligned to a sequence L of N characters from some finite alphabet. Let $t_i$ be the position in L where $S_i$ starts. An alignment is called feasible threading if:

1) $t_i \geq t_{i-1} + l_{i-1}$ for all i,

2) the length $g_i$ (called gap or loop) of uncovered characters; i.e $g_i = t_i - t_{i-1} - l_{i-1}$ is bounded, say $\min g_i \leq g_i \leq \max g_i$.

Each feasible threading $t = (t_1, t_2, .., t_m)$ is scored by a function $f(t) = \sum f_i(t_i) + \sum h_i(g_i)$ where $f_i$ score the placement of the segment i to a given position $t_i$ and $h_i$ is used in some experiments for scoring the gap between two consecutive segments. If the problem now is to minimize f(t) over the set F of feasible threading, one can show the equivalents with the shortest path problem between two vertices of a very structured graph.

The model of protein threading problem is to minimize the objective function

$$\sum_i \sum_j g_{ij}\, x_{ij} + \sum_{(i,k)\in E} \sum_{j\leq l} \bar{g}_{ijkl}\, x_{ij} x_{kl}$$

Subject to

$$\sum_{j=1}^{n} x_{ij} = 1; \quad i = \overline{1,m}$$

$$x_{ij} \leq \sum_{k=1}^{j} x_{i-1,k}, \quad i = \overline{2,m},\ j = \overline{1,n-1}$$

Where m is the number of segments, $n = N - \sum_{k=1}^{m} l_k + 1$ (The number $l_k$ are the lengths of the segments increased by $l_k^{min}$ the minimal number of gaps between the segments k and k+1) is the number of possible placements of each segment relative to the end of the previous one, $x_{ij}$ are binary variables with $x_{ij} = 1$ meaning the segment i starts from the obuolute position $\sum_{k=1}^{i-1} l_k + j$ of the position sequence L.

Many different algorithms have been proposed for finding the correct threading of a sequence onto a structure, though many make use of dynamic programming in some form. For full 3-D threading, the problem of identifying the best alignment is very difficult (it is an NP-hard problem). Researchers have made use of many combinatorial optimization methods to arrive at solutions. There are many algorithms, for example, the protein threading software RAPTOR, which is based on linear integer programming.

In this paper, we focus on developing efficient algorithms. We notice that the mathematical models used in the literatures are normally a linear integer programming, which can actually be regarded as a linearization of a quadratic integer programming problem. This motivates us to study the original quadratic integer programming directly. Recently, quadratic integer programming becomes a hot research topic in optimization society. Many mathematical tools such as conic programming are developed, with which we can construct corresponding efficient algorithms.

Now, consider the zero-one quadratic programming problem

$$P: \min C^T x + x^T Q x \quad (1,1)$$
$$s.t. \ h^T x + x^T G x \geq g \quad (1,2)$$
$$x \in X \subseteq \{0,1\}, \quad (1,3)$$

where Q and G are general symmetric matrices of dimension $n \times n$.

This problem is a generalization of unconstrained zero-one quadratic problems, zero-one quadratic knapsack problems, quadratic assignment problems and so on. It is clearly NP-hard.

Linearization strategies are to reformulate the zero-one quadratic programs as equivalent mixed-integer programming problems (1.1) and (1.3) with additional binary variables and/or continuous variables and continuous constraints, see [1, 2, 3, 6, 7, 8, 9, 10, 12, 13].

Recently, Sherali and Smith [14] developed small linearizations for (1.1) - (1.3), which is more general with structure. The linearization generated by our approach is smaller. More tight linearization strategies are proposed in this article for further improvement.

This article is organized as follows. In section 2, we shortly describe the existing efficient linearization approach. In section 3, we introduce our approach and represent the linearized model. We conclude the paper in section 4.

## 2 THE EXISTING EFFICIENT LINEARIZATION APPROACH

Define

$$\gamma^i_{min/max} = \min/\max\{Q_i x : x \in \bar{X}\}, \forall i, \quad (2.1)$$

where $Q_i$ is the i-th row of Q, and $\bar{X}$ is any suitable relaxation of X such that the problem (2.1) can be solved relatively easily. $\gamma_{min/max}$ be the vector with components $\gamma^i_{min/max}$, $i = 1, 2, ..., n$, and

$$\Gamma_{min/max} = diag(\gamma^i_{min/max}). \text{ Similarly, define}$$

$$\lambda^i_{min/max} = \min/\max\{G_i x : x \in \bar{X}\}, \forall i, \quad (2.2)$$

and

$$\lambda_{min/max} = (\lambda^i_{min/max}, i=1,...,n)^T,$$
$$\Lambda_{min/max} = diag(\lambda^i_{min/max}, i=1,...,n).$$

Sherali and Smith [14] reformulated Problem P as an equivalent bilinearly constrained bilinear problem by introducing $\gamma = Qx$ and $\lambda Gx$. Linearizing the terms $x_i \gamma_i$ and $x_i \lambda_i$ by $s'_i$ and $z'_i$ respectively, they obtained

$$BP : \min c^T x + e^T S' \quad (2,3)$$
$$s.t. \ Qx = \gamma \quad (2,4)$$
$$h^T x + e^T z' \geq g \quad (2,5)$$
$$Gx = \lambda \quad (2,6)$$
$$\gamma^i_{min} x_i \leq s'_i \leq \gamma^i_{max} x_i, \forall i \quad (2,7)$$
$$\gamma_{min} x_i \leq S'_i \leq \gamma_{max} x_i (1-x_i), \forall i, \quad (2,8)$$
$$\lambda^i_{min} x_i \leq z'_i \leq \lambda^i_{max} x_i, \forall i, \quad (2,9)$$
$$\lambda i_{min}(1-x_i) \leq (\lambda_i - z'_i) \leq \lambda^i_{max}(1-x_i), \forall i, \quad (2,10)$$
$$x \in X \quad (2,11)$$

where e is a conformable vector of ones and the constrains (2.7) - (2.10) comes from multiplying

$$\gamma_{\min} \leq \gamma \leq \gamma_{\max}, \quad \lambda_{\min} \leq \lambda \leq \lambda_{\max} \quad (2.12)$$

by $x_i$ and $(1-x_i)$.

BP (2.3) - (2.11) has the following equivalent compact formulation

$$\text{BP:} \quad \min \ c^T x + e^T s + \gamma_{\min}^T x \quad (2.13)$$
$$\text{s.t.} \quad Qx = y + s + \Gamma_{\min} e \quad (2.14)$$
$$h^T x + e^T z + \lambda_{\min}^T x \geq g \quad (2.15)$$
$$Gx = \lambda \quad (2.16)$$
$$0 \leq s_i \leq (\gamma_{\max}^i - \gamma_{\min}^i) x_i, \forall i, \quad (2.17)$$
$$0 \leq y_i \leq (\gamma_{\max}^i - \gamma_{\min}^i)(1 - x_i), \forall i, \quad (2.18)$$
$$0 \leq z_i \leq (\lambda_{\max}^i - \lambda_{\min}^i) x_i \quad (2.19)$$
$$\lambda_{\min}^i \leq (\lambda_i - z_i) \leq \lambda_{\max}^i - (\lambda_{\max}^i - \lambda_{\min}^i) x_i, \forall i, \quad (2.20)$$
$$x \in X \quad (2.21)$$

via the linear transformation

$$\begin{aligned} s_i &= s_i' - \gamma_{\min}^i x_i, \forall i, \\ y_i &= \gamma_i - s_i' - \gamma_{\min}^i (1 - x_i), \forall i, \quad (2.22) \\ z_i &= z_i' - \lambda_{\min}^i x_i, \forall i, \end{aligned}$$

Since the optimization and constraint senses of BP tend to push the variables s to their lower bounds and z to their upper bounds, the final relaxed version of BP was written as

$$\overline{\text{BP}}: \quad \min \ c^T x + e^T s + \gamma_{\min}^T x \quad (2.23)$$
$$\text{s.t.} \quad Qx = y + s + \Gamma_{\min} e \quad (2.24)$$
$$0 \leq y \leq [\Gamma_{\max} - \Gamma_{\min}](e - x) \quad (2.25)$$
$$s \geq 0 \quad (2.26)$$
$$h^T x + e^T z + \lambda_{\min}^T x \geq g \quad (2.27)$$
$$Gx \geq z + \lambda_{\min} \quad (2.28)$$
$$0 \leq z \leq [\Lambda_{\max} - \Lambda_{\min}] x \quad (2.29)$$
$$x \in X, \quad (2.30)$$

by deleting the upper bounding inequalities for $s$ and $\lambda - z$ in (2.17) and (2.20), and combining (2.16) with (2.20).

It was shown in [14] that Problems *BP* and *P* are equivalent in the sense that for each feasible solution to one problem, there exists a feasible solution to the other problem having the same objective value. Furthermore, let $x$ be part of an optimal solution to Problem *BP*. Then $x$ solves Problem *P*.

Besides, *BP* can be improved by the additional cuts

$$(\lambda_{\min}^i - \gamma_{\min}^i - w_{\max}^j) x_i - s_i + z_i \leq 0, \forall i, \quad (2.31)$$

which is derived from multiplying $\lambda_i - \gamma_i \leq w_{\max}^i$ by $x_i$ where $w_{\max}^i = \max\{(G_i - Q_i)x : x \in \overline{X}\}$.

## 3 A REPRESENTATION APPROACH

Motivated by [15], we first reveal the relation between general quadratic and piece-wise linear terms for zero-one variables.

**Lemma 3.1.** Let $x \in X \subseteq \{0,1\}^n$. for all $i = 1, ..., n$,

$$x_i Q_i x = \max\{\gamma_{\min}^i x_i, Q_i x + \gamma_{\max}^i x_i - \gamma_{\max}^i\}, \quad (3.1)$$
$$x_i Q_i x = \min\{\gamma_{\max}^i x_i, Q_i x + \gamma_{\min}^i x_i - \gamma_{\min}^i\} \quad (3.2)$$

*Proof.* Suppose $x_i = 0$, the left hand side of (3.1) is clearly 0 and the right hand side becomes $\max\{0, Q_i x - \gamma_{\max}^i\} = 0$. On the other hand, if $x_i \neq 0$, it must hold that $x_i = 1$, the right hand side of (3.1) reads $\max\{\gamma_{\max}^i, Q_i x\} = Q_i x$, which is equal to the left hand side. The proof of (3.1) is completed and (3.2) can be similarly verified.

**Corollary 3.1.**

Let $x \in X \subseteq \{0,1\}^n$. for all $i = 1, ..., n$,

$$\max\{\gamma_{\min}^i x_i, Q_i x + \gamma_{\max}^i x_i - \gamma_{\max}^i\} \leq s_i' \leq \quad (3.3)$$
$$\leq \min\{\gamma_{\max}^i x_i, Q_i x + \gamma_{\min}^i x_i - \gamma_{\min}^i\},$$

if and only if

$$s_i' = x_i Q_i x. \quad (3.4)$$

Combining (3.1) with (3.2), we have

$$\begin{aligned} x_i Q_i x &= \max\{\gamma_{\min}^i x_i, Q_i x + \gamma_{\max}^i x_i - \gamma_{\max}^i\} \quad (3.5) \\ &= \min\{\gamma_{\max}^i x_i, Q_i x + \gamma_{\min}^i x_i - \gamma_{\min}^i\}, \end{aligned}$$

The above results hold true for $G_i$ and $\lambda_i$ defined before. Linearization based on Corollary 3.1 is just BP (2.3) - (2.11), where the linear inequalities (2.7) - (2.8) is nothing but (3.3). We remark here the four inequalities implied by (3.3) were first introduced in [8]. Actually, not all inequalities (3,3) are necessary in the final linearized model. To see this, below we first introduce the principle of reformulating zero-one quadratic programs into piece-wise linear programs. Generally, for continuous programs, we have

**Proposition 3.1**. Any convex program with linear or piece-wise linear objective function and constraints is equivalent to a linear program in the sense that there is a one-to-one projection between both feasible solutions.

*Proof.* We notice that $\min f(x)$ is equivalent to

$\min t$

$s.t.\ t - f(x) \geq 0$

Without loss of generality we assume that the objective function is linear. The constraint set is convex and characterized by piece-wise linear inequalities. It follows that it is convex polyhedral, which must have linear expression.

It is easy to see that the equivalence of Proposition 3.1 holds if we restrict the variables to be zeros or ones. Next we show the existence of such equivalent 'convex' piece-wise linear program for zero-one quadratic minimization problem.

**Proposition 3.2.** For any zero-one quadratic minimization problem, there is an equivalent zero-one piece-wise linear program with convex objective function and constraints.

Proof. Clearly, the maximum of several linear functions is convex and the minimum is concave. Then (3.1) and (3.2) in Lemma 3.1 provide the convex and concave formulations, respectively. Therefore, for any given zero-one quadratic minimization problem, we can obtain an equivalent convex piece-wise linear program by using (3.1) and/or (3.2). Note that we use (3.1) and (3.2) simultaneously only when handling equality constraints, see also Corollary 3.1.

Now we can see that (1.1) - (1.3) has the following equivalent formulation

$$\min\ c^T x + \sum_{i=1}^{n} \max\{\gamma_{\min}^i x_i, Q_i x + \gamma_{\max}^i x_i - \gamma_{\max}^i\} \quad (3.6)$$

$$s.t.\ h^T x + \sum_{i=1}^{n} \min\{\lambda_{\max}^i x_i, G_i x + \lambda_{\min}^i x_i - \lambda_{\min}^i\} \geq g, \quad (3.7)$$

$$x \in X \subseteq \{0,1\}^n. \quad (3.8)$$

Linearizing (3.6)-(3.8) becomes very easy. For example, (3.7) is equivalent to

$$h^T x + \sum_{i=1}^{n} z_i \geq g, \quad (3.9)$$

$$z_i \leq \lambda_{\max}^i x_i, \quad (3.10)$$

$$z_i \leq G_i x + \lambda_{\min}^i x_i - \lambda_{\min}^i, \quad (3.11)$$

since (3.9)-(3.11) is a relaxation of (3.7) and (3.9)-(3.11) also implies (3.7).

Now we can obtain a linearization for (3.6)-(3.8), which is similarly to $\overline{BP}$ except that we do not require $y \geq 0$ and $z \geq 0$. In other words, they are redundant in $\overline{BP}$.

Finally, we point out that the non-necessity of inequalities such as $y \geq 0$ and $z \geq 0$ was also observed in [1, 2]. Actually, the linearization generated by our convex piece-wise approach coincides theirs.

## 4   CONCLUSIONS

In this article, we defined the protein folding problem and discussed its solution through presenting small linearizations for the zero-one quadratic minimization problem.

We present the equivalence of quadratic terms and piece-wise linear terms for zero-one variables. There are two piece-wise formulations, convex and concave cases. We show the smaller linearization is based on the convex piece-wise objective function and constraints. Linearization generated by our approach is smaller than that in [14]. Our approach can be easily extended to linearize polynomial zero-one minimization problems which have many applications, particularly in biological computing problems.